\documentclass[10pt,prl,twocolumn,superscriptaddress]{revtex4-1}
\usepackage{graphicx,amsmath,amssymb}
\usepackage{epsfig}
\usepackage[colorlinks=true,citecolor=blue,
linkcolor=blue,urlcolor=blue]{hyperref}
\begin{document}

\renewcommand{\Re}{\mathop{\mathrm{Re}}}
\renewcommand{\Im}{\mathop{\mathrm{Im}}}
\renewcommand{\b}[1]{\mathbf{#1}}
\renewcommand{\u}{\uparrow}
\renewcommand{\d}{\downarrow}
\newcommand{\bsigma}{\boldsymbol{\sigma}}
\newcommand{\blambda}{\boldsymbol{\lambda}}
\newcommand{\Tr}{\mathop{\mathrm{Tr}}}
\newcommand{\sgn}{\mathop{\mathrm{sgn}}}
\newcommand{\sech}{\mathop{\mathrm{sech}}}
\newcommand{\diag}{\mathop{\mathrm{diag}}}
\newcommand{\half}{{\textstyle\frac{1}{2}}}
\newcommand{\sh}{{\textstyle{\frac{1}{2}}}}
\newcommand{\ish}{{\textstyle{\frac{i}{2}}}}
\newcommand{\thf}{{\textstyle{\frac{3}{2}}}}
\newcommand{\be}{\begin{equation}}
\newcommand{\ee}{\end{equation}}

\title{Topological order in a correlated three-dimensional topological insulator}

\author{Joseph Maciejko}
\email[electronic address: ]{jmaciejk@princeton.edu}
\affiliation{Princeton Center for Theoretical Science, Princeton University, Princeton, New Jersey 08544, USA}

\author{Victor Chua}
\affiliation{Department of Physics, The University of Texas at Austin, Austin, Texas 78712, USA}

\author{Gregory A. Fiete}
\affiliation{Department of Physics, The University of Texas at Austin, Austin, Texas 78712, USA}

\date\today

\begin{abstract}
Motivated by experimental progress in the growth of heavy transition metal oxides, we theoretically study a class of lattice models of interacting fermions with strong spin-orbit coupling. Focusing on interactions of intermediate strength, we derive a low-energy effective field theory for a fully gapped, topologically ordered, fractionalized state with an eight-fold ground-state degeneracy. This state is a fermionic symmetry-enriched topological phase with particle-number conservation and time-reversal symmetry. The topological terms in the effective field theory describe a quantized magnetoelectric response and nontrivial mutual braiding statistics of dynamical extended vortex loops with emergent fermions in the bulk. We explicitly compute the expected mutual statistics in a specific model on the pyrochlore lattice within a slave-particle mean-field theory. We argue that our model also provides a possible condensed-matter realization of oblique confinement.
\end{abstract}

\maketitle

The study of three-dimensional topological insulators (TI) has become one of the most active areas of research in condensed matter physics~\cite{hasan2010,qi2011}. Their striking properties, such as robust surface states observed by angle-resolved photoemission spectroscopy~\cite{hsieh2008} and a topological magnetoelectric response~\cite{qi2008} leading to the prediction of a quantized Kerr and Faraday effect~\cite{tse2010,maciejko2010}, can be understood in a single-particle picture where electron-electron interactions play no significant role.

A burning question in TI research is how interactions beyond the perturbative limit affect their properties~\cite{raghu2008,zhang2009,
hohenadler2013,witczak-krempa2013}. If the interaction strength is much less than the bulk energy gap, the bulk remains adiabatically connected to a TI and the problem reduces to studying the effect of interactions on the surface states. On the other hand, if the interaction strength is comparable to or greater than the bulk energy gap, the TI phase may disappear altogether and the problem has been comparatively less studied. Recent work~\cite{pesin2010,kargarian2011} suggests the intriguing possibility that for sufficiently strong interactions the TI phase might give way to an exotic phase dubbed the topological Mott insulator (TMI)---or its topological-crystalline-insulator (TCI) cousin the TCMI \cite{Kargarian:prl13}, which can be understood as a TI or TCI of charge-neutral, fermionic spinons interacting with gapless ``photon'' excitations of a deconfined $U(1)$ gauge field. The TMI is thus essentially a 3D quantum spin liquid~\cite{wen2002} with no charge response, albeit with a peculiar band structure for the spinons~\cite{fiete2012}, and has indeed been proposed as a possible ground state for a pure spin model on the pyrochlore lattice~\cite{bhattacharjee2012}.

In this paper we investigate the possibility of a new state of matter---denoted by TI* in the following---which is intermediate between the TI and the TMI, in the sense that it is not adiabatically connected to the TI yet exhibits a nontrivial charge response unlike the TMI. Unlike the TI and the TMI, the TI* is a topologically ordered, fully gapped state with eight degenerate ground states on the three-torus $T^3$ (corresponding to periodic boundary conditions in all three directions of space). It can be understood as a TI of emergent fermionic excitations that are electrically charged, yet not adiabatically connected to the microscopic electrons. These fermionic excitations interact with an emergent, deconfined $\mathbb{Z}_2$ gauge field that is ultimately responsible for the topological order.


We are interested in a class of electron lattice models of the form
\begin{align}\label{H}
H=\sum_{rr'}\sum_{\alpha\beta}t^{rr'}_{\alpha\beta}c_{r\alpha}^\dag
c_{r'\beta}+\frac{U}{2}\sum_r\left(\sum_\alpha n_{r\alpha}-1\right)^2,
\end{align}
at half-filling $\langle\sum_\alpha n_{r\alpha}\rangle=1$, where $c_{r\alpha}^\dag$ ($c_{r\alpha}$) is a creation (annihilation) operator for an electron of spin $\alpha=\uparrow,\downarrow$ at site $r$, $n_{r\alpha}=c_{r\alpha}^\dag c_{r\alpha}$ is the number of electrons of spin $\alpha$ on site $r$, $t^{rr'}_{\alpha\beta}$ is a spin-dependent hopping amplitude and $U>0$ is the on-site Hubbard interaction. Although we will later present numerical results for a specific model, as far as the universal properties of the TI* are concerned we only require that Eq.~(\ref{H}) correspond to an ordinary TI in the noninteracting limit $U=0$. In the $U\rightarrow\infty$ limit, the system is described by a spin-$\frac{1}{2}$ Heisenberg-type model which is expected to have a magnetic ground state~\cite{pesin2010,witczak-krempa2012}.

In order to study the possible phases of (\ref{H}) at intermediate values of $U$, we make use of the recently introduced $\mathbb{Z}_2$ slave-spin theory for correlated electron systems~\cite{huber2009,ruegg2010,nandkishore2012,
ruegg2012,maciejko2013}. This theory is based on the simple observation that the Hubbard interaction energy in (\ref{H}) depends only on the total occupation of site $r$ \emph{modulo 2}, which can be represented by a Pauli matrix (Ising variable) $\tau^z_r$, viz. $\tau^z_r=1$ for unoccupied or doubly occupied sites and $\tau^z_r=-1$ for singly occupied sites. The annihilation/creation of an electron at site $r$ changes this occupation modulo 2, and the electron annihilation (creation) operator is written as $c_{r\alpha}^{(\dag)}=f_{r\alpha}^{(\dag)}\tau^x_r$ where the slave-spin $\tau^x_r$ flips the sign of $\tau^z_r$. The slave-fermion $f_{r\alpha}^{(\dag)}$ carries the same spin and charge quantum numbers as the electron. The Hamiltonian (\ref{H}) is written in terms of the slave-spins and slave-fermions as~\cite{Supplemental}
\begin{align}\label{Hss}
H=\sum_{rr'}\sum_{\alpha\beta}t^{rr'}_{\alpha\beta}\tau^x_r\tau^x_{r'}
f_{r\alpha}^\dag f_{r'\beta}
+\frac{U}{4}\sum_r\left(\tau^z_r+1\right).
\end{align}
While (\ref{H}) acts in the physical Hilbert space of electrons, (\ref{Hss}) acts in the enlarged Hilbert space of slave-spins and slave-fermions. To project back onto the physical Hilbert space we impose the local constraint $G_r=1$ on each site $r$ where the unitary operator $G_r=(-1)^{\sum_\alpha f^\dag_{r\alpha}f_{r\alpha}+\frac{1}{2}(\tau^z_r-1)}$, which performs $\mathbb{Z}_2$ gauge transformations $f_{r\alpha}^{(\dag)}\rightarrow-f_{r\alpha}^{(\dag)}$, $\tau^x_r\rightarrow-\tau^x_r$, commutes with the Hamiltonian (\ref{Hss}). An effective way to implement this local constraint is to pass to a path integral representation~\cite{senthil2000}. Defining the partition function by $Z=\Tr(e^{-\beta H}P)$ where $P=\prod_r[(1+G_r)/2]$ is a projector and $\beta$ is the inverse temperature ensures that only physical states are included in the partition sum. $Z$ can be mapped in this way to the partition function of a $\mathbb{Z}_2$ gauge theory in 4D Euclidean spacetime with bosonic and fermionic matter in the fundamental representation~\cite{Supplemental},
\begin{align}\label{Zgauge}
Z=\int D\bar{f}_{i\alpha}Df_{i\alpha}\sum_{\{\tau^x_i\}}
\sum_{\{\sigma_{ij}\}}e^{-S_{\mathbb{Z}_2}[\bar{f},f,\tau^x,\sigma]},
\end{align}
where the action is given by $S_{\mathbb{Z}_2}=S_{\tau^x}+S_f+S_B$, with
\begin{align}
S_{\tau^x}&=-\kappa\sum_{ij}\tau^x_i\sigma_{ij}\tau^x_j,\\
S_f&=-\sum_{ij}\sum_{\alpha\beta} t^{ij}_{\alpha\beta}\bar{f}_{i\alpha}\sigma_{ij}f_{j\beta},\\
e^{-S_B}&=\prod_{i,j=i-\hat{\tau}}\sigma_{ij},
\end{align}
where $\sigma_{ij}=\pm 1$ is a spacetime $\mathbb{Z}_2$ gauge field, $i,j$ are sites on a 4D spacetime lattice, and $t^{ij}_{\alpha\beta}$ is proportional to $t^{rr'}_{\alpha\beta}$ on spatial links and equal to $-1$ on temporal links. $S_B$ is a Berry phase term~\cite{senthil2000} which corresponds to a background $\mathbb{Z}_2$ charge on every site~\cite{moessner2001c}, and $\kappa$ is a constant which depends on the Hubbard interaction $U$. The effective gauge theory (\ref{Zgauge}) reproduces the limiting cases mentioned earlier. The noninteracting limit $U=0$ corresponds to $\kappa\rightarrow\infty$ where the gauge fields are frozen and the slave-spins are ferromagnetically ordered $\tau^x_r=1$ (or gauge equivalent configurations)~\cite{fradkin1979}, such that the slave-fermions are identified with the original electrons. In the $U\rightarrow\infty$ limit we have $\kappa=0$. Integrating out the slave-spins and gauge field generates interactions between slave-fermions.  The Berry phase term imposes the constraint of one slave-fermion per site, which corresponds to a spin-$\frac{1}{2}$ Heisenberg-type model as expected~\cite{Supplemental}.

We now proceed to derive the low-energy effective field theory of the TI* phase. In order to describe the electromagnetic response of the TI*, which involves the $U(1)$ electromagnetic gauge field $A_\mu$, it is more convenient to work with $U(1)$ gauge fields than $\mathbb{Z}_2$ gauge fields. When a charge-2 scalar field coupled to a $U(1)$ gauge field Bose condenses, the $U(1)$ gauge structure is broken down to a $\mathbb{Z}_2$ gauge structure, as is familiar from the condensation of charge-2$e$ Cooper pairs in superconductivity~\cite{hansson2004}. Reversing this argument, the $\mathbb{Z}_2$ gauge theory (\ref{Zgauge}) can be written as the theory of a $U(1)$ gauge field $a_{ij}$ coupled to an extra scalar field $n_{ij}$ defined on the links of the spacetime lattice~\cite{Supplemental}. For large but finite $U$, $\kappa$ is small and the slave-spins are gapped. The slave-fermions inherit the band structure of the noninteracting TI Hamiltonian and are also gapped. Therefore, we expect that the resulting theory has a gapped deconfined phase~\cite{wegner1971} for large but finite values of $U$. In what follows we focus on this deconfined phase which we denote the TI* phase.

The effective field theory of the TI* phase is obtained by integrating out all gapped matter fields: the slave-spins and slave-fermions. According to Ref.~\onlinecite{senthil2000}, the Berry phase term $S_B$ enhances the stability of the deconfined phase but does not otherwise affect its long-wavelength, low-energy properties, which are the focus of our analysis. We will drop $S_B$ in what follows. Because the TI* phase is a deconfined phase, monopoles are confined and we can take the continuum limit $a_{ij}\rightarrow a_\mu$. Integrating out the slave-spins generates non-topological terms for $a_\mu$ such as the Maxwell term~\cite{fradkin1979}. The coupling between the $U(1)$ gauge field $a_\mu$ and the charge-2 link variable $n_{ij}\rightarrow n_\mu$ becomes a level-2 $BF$ term~\cite{blau1991,bergeron1995,szabo1998,
hansson2004} in $(3+1)$ dimensions~\cite{Supplemental}. $BF$ terms with different levels and physical interpretations have appeared recently in effective field theories of TIs~\cite{cho2011,diamantini2012,chan2013}. The slave-fermions couple to the external electromagnetic gauge field $A_\mu$ with charge $e$ as well as to the internal gauge field $a_\mu$ (with charge 1). Because of their topological band structure, integrating out the slave-fermions generates a $\theta$-term~\cite{qi2008} for the combined gauge field $a_\mu+eA_\mu$. Ignoring all non-topological terms, the universal properties of the TI* phase are encoded in its topological field theory,
\begin{align}\label{LTI}
\mathcal{L}_{\textrm{TI*}}=\frac{p}{4\pi}\epsilon^{\mu\nu\lambda\rho}b_{\mu\nu}\partial_\lambda (a_\rho-eA_\rho)+\frac{\theta}{32\pi^2}\epsilon^{\mu\nu\lambda\rho}f_{\mu\nu}f_{\lambda\rho},
\end{align}
the main result of this work, where $f_{\mu\nu}=\partial_\mu a_\nu-\partial_\nu a_\mu$ is the field strength of $a_\mu$, $\theta=(2m+1)\pi$, $m\in\mathbb{Z}$ is the axion angle of the noninteracting TI Hamiltonian, and $p=2$. The Lagrangian (\ref{LTI}) with $p=1$ was obtained by Chan {\it et al.}~\cite{chan2013,BFnormalization} as the effective theory of a noninteracting TI, and upon quantization correctly produces~\cite{bergeron1995} a unique ground state on $T^3$. It was observed in Ref.~\cite{cho2011,chan2013} that a $BF$ term with level $p>1$ would describe a fractionalized TI with multiple ground states on $T^3$. Here we show that a level-2 $BF$ term arises naturally for a TI subject to a strong on-site Hubbard interaction $U$ due to the emergent $\mathbb{Z}_2$ gauge structure of the partition function (\ref{Zgauge}). As a result, Eq.~(\ref{LTI}) predicts a ground-state degeneracy of $2^3=8$ on $T^3$~\cite{bergeron1995,hansson2004} for the TI* phase.

The universal electromagnetic response of the TI* phase can also be extracted from Eq.~(\ref{LTI}). Integrating out $b_{\mu\nu}$ which sets $f_{\mu\nu}=eF_{\mu\nu}$ with $F_{\mu\nu}=\partial_\mu A_\nu-\partial_\nu A_\mu$ the electromagnetic field strength, we obtain
\begin{align}
\mathcal{L}_\textrm{em}=\frac{\theta e^2}{32\pi^2}\epsilon^{\mu\nu\lambda\rho}F_{\mu\nu}F_{\lambda\rho},
\end{align}
i.e., the topological magnetoelectric effect~\cite{qi2008} and associated surface magnetooptical effects~\cite{tse2010,maciejko2010} are the same as for the noninteracting TI, at least for weak electromagnetic fields. This contrasts with the TMI~\cite{pesin2010}, which has no quantized electromagnetic response. Another difference with the TMI is that the TI* is a fully gapped topological phase with no gapless ``photon'' excitations. The (gapped) excitations of the TI* are the slave-spins, the slave-fermions, and $\mathbb{Z}_2$ vortex loops (closed $\pi$ flux tubes) from the gauge sector~\cite{wegner1971}. One might expect the vortex loops to carry a gapless (1+1)D helical liquid of slave-fermions~\cite{ran2008,rosenberg2010b,ostrovsky2010,chiu2011}, but since the vortex loops have a finite creation energy per unit length the fermionic modes, while remaining Kramers degenerate, have a discrete energy spectrum with finite-size gaps $\propto 1/L$ where $L$ is the length of the loop. The vortex loops and slave-particles (slave-spins and slave-fermions) are described in the effective field theory (\ref{LTI}) by a conserved string current $\Sigma^{\mu\nu}$ and particle current $j^\mu$, respectively, that couple to the gauge fields as $\frac{1}{2}\Sigma^{\mu\nu}b_{\mu\nu}$ and $j^\mu a_\mu$~\cite{bergeron1995}. The level-2 $BF$ term then implies a statistical phase of $\frac{2\pi}{p}=\pi$ for braiding a slave-particle around a vortex loop~\cite{bergeron1995,hansson2004}.

$\mathbb{Z}_2$ gauge theory in (3+1)D exhibits a confinement-deconfinement transition at zero temperature~\cite{wegner1971}. Given that the TI* corresponds to the deconfined phase of this theory, it is natural to ask what happens when the $\mathbb{Z}_2$ gauge field enters a confined phase. For the TMI, the confined phase of the $U(1)$ gauge field is a monopole condensate~\cite{cho2012}, with the added interesting feature that the monopole acquires a $U(1)$ charge via the Witten effect~\cite{witten1979,qi2008,rosenberg2010} arising from the topological band structure of the spinons. For the TI*, the $\mathbb{Z}_2$ nature of the gauge theory combined with the topological band structure of the slave-fermions implies the possibility of an exotic type of confinement known as \emph{oblique confinement}, first proposed by 't Hooft~\cite{thooft1981} in the context of gauge theories of particle physics and shown to occur in $\mathbb{Z}_p$ gauge theories~\cite{cardy1982a,cardy1982b}. A more detailed discussion of this oblique confined phase will be reported in a separate publication.

\begin{figure}
\includegraphics[width=\columnwidth]{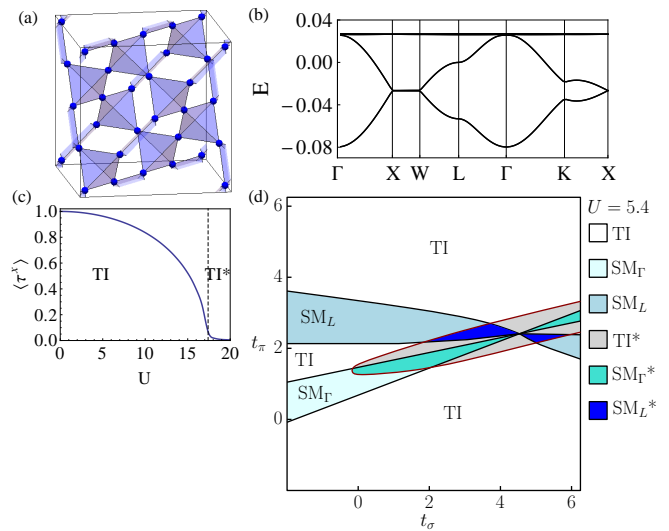}
\caption{(color online) (a) Pyrochlore lattice of corner-sharing tetrahedra. (b) Electronic band structure near the multicritical point $(t_{\sigma c},t_{\pi c})\approx (4.62,2.43)$ at $U=0$. The mean-field slave-fermion band structure is related by a uniform rescaling. The Fermi energy lies between the dispersive and flat bands. (c) $\langle \tau^x \rangle$ as a function of $U$ at $(t_\sigma,t_\pi)=(0,0)$. The TI-TI* transition occurs at $U\approx 17$. (d) Slave-particle mean-field phase diagram in the $(t_\sigma,t_\pi)$ plane at $U=5.4$. The semi-metallic phases SM$_\Gamma$ and SM$_L$ are distinguished by their Fermi points $\Gamma$ and $L$. The red line encloses the $\mathbb{Z}_2$ fractionalized phases (darker shade). Increasing $U$ expands these phases which emanate from the multicritical point.}\label{PD}
\end{figure}

To provide a concrete example of the TI* phase, we consider a particular instance of Eq.~(\ref{H}), a fermion model on the pyrochlore lattice [Fig.~\ref{PD}(a)]. This model~\cite{pesin2010,kargarian2011,witczak-krempa2012,Kargarian:prl13} is partially motivated by the A$_2$Ir$_2$O$_7$ pyrochlore iridates (A is a rare earth element), materials that exhibit both strong correlations and strong spin-orbit coupling~\cite{witczak-krempa2013}. However, for the purpose of exploring the feasibility of the TI* phase we could also consider any other 3D lattice model whose non-interacting limit ($U=0$) is a TI, such as models previously considered on the diamond~\cite{fu2007} or perovskite~\cite{weeks2010} lattices. The noninteracting Hamiltonian matrix $t^{rr'}_{\alpha\beta}$ in Eq.~(\ref{H}) is parametrized by the hopping amplitudes $t_\sigma$ and $t_\pi$~\cite{witczak-krempa2012}.

We solve the slave-particle Hamiltonian (\ref{Hss}) within the mean-field or saddle-point approximation by decoupling the kinetic term and introducing mean-fields $\chi^{rr'}_{\alpha\beta}$ and $J_{rr'}$ that obey the self-consistent equations
\begin{align}
\chi^{rr'}_{\alpha\beta} &= t^{rr'}_{\alpha\beta}\sigma_{rr'} \langle 
\tau_r^x \tau_{r'}^x \rangle_\textrm{MF}, \label{Chimf}\\
J_{rr'} &=  \sum_{\alpha\beta} t^{rr'}_{\alpha\beta}\sigma_{rr'} \langle f_{r\alpha}^\dagger f_{r'\beta} \rangle_\textrm{MF},\label{Jmf}
\end{align}
where $\langle\cdots\rangle_\textrm{MF}$ denotes an expectation value in the ground state of the mean-field Hamiltonian $H_\textrm{MF} = \sum_{rr'}\sum_{\alpha\beta} \chi^{rr'}_{\alpha\beta}\sigma_{rr'} f_{r\alpha}^\dagger f_{r'\beta}+\sum_{rr'} J_{rr'} \sigma_{rr'}\tau_r^x \tau_{r'}^x + \frac{U}{4}\sum_{r}(\tau_r^z+1)$.  The fields $\sigma_{rr'}=\pm 1$ are the spatial components of the $\mathbb{Z}_2$ gauge fields $\sigma_{ij}$ in Eq.~(\ref{Zgauge}). Their inclusion in $H_\textrm{MF}$ is required to maintain the $\mathbb{Z}_2$ gauge symmetry which is otherwise broken by the mean-field approximation. $H_\textrm{MF}$ describes free slave-fermions $f,f^\dag$ propagating in a static background $\mathbb{Z}_2$ gauge field and an interacting transverse-field Ising model (TIM) in the slave-spins $\tau^x$. The $\tau^x$ sector is solved in the cluster approximation~\cite{ruegg2012} where 16 interacting slave-spins are coupled to a mean field $\langle \tau^x \rangle$ that is solved for self-consistently. Solving Eq.~(\ref{Chimf}-\ref{Jmf}) numerically, we find that a uniform $\mathbb{Z}_2$ gauge field $\sigma_{rr'}=+1$ and the uniform ferromagnetic ansatz $J_{rr'}=J<0$ have the least variational energy. Fixing the amplitudes of $\chi^{rr'}_{\alpha\beta}$ and $J_{rr'}$ to be constant in the mean-field Hamiltonian approach corresponds to the constant-amplitude approximation in the path-integral approach~\cite{Supplemental}. 

For small $U$, the slave-spins order $\langle \tau_r^x \rangle\neq 0$ and $c_{r\alpha}^{(\dagger)}$ is proportional to $f_{r\alpha}^{(\dagger)}$, implying that the weakly interacting quasiparticles inherit the properties of the $U=0$ phases to which they are adiabatically connected. The $U=0$ limit of the model considered has a strong TI phase and two distinct semi-metallic (SM) gapless phases~\cite{witczak-krempa2012} with Fermi points at the $\Gamma$ and $L$ inversion-symmetric points. These phases survive at finite $U$ [Fig.~\ref{PD}(d)]. For sufficiently large $U$ however, the TIM enters the disordered phase $\langle \tau^x_r \rangle =0$ and slave-spin excitations are gapped. The deconfined TI* phase is identified with the phase where the slave-fermion sector maintains a gapped TI band structure. If the latter is semi-metallic, we obtain a new gapless phase denoted SM* which is not adiabatically connected to a weakly correlated SM. Without direct hoppings $t_\sigma$ and $t_\pi$, the TI* and SM* phases require strong interactions [Fig.~\ref{PD}(c)]. However, they can appear at relatively modest $U$ if $(t_\sigma,t_\pi)$ is tuned to the vicinity of a multicritical point where the SM-TI phase boundaries intersect [Fig.~\ref{PD}(d),(b)]. At this multicritical point the noninteracting band structure [Fig.~\ref{PD}(b)] acquires the dispersion of the nearest-neighbor tight-binding model on the pyrochlore lattice with real isotropic hoppings~\cite{guo2009}. 

\begin{figure}
\includegraphics[width=\columnwidth]{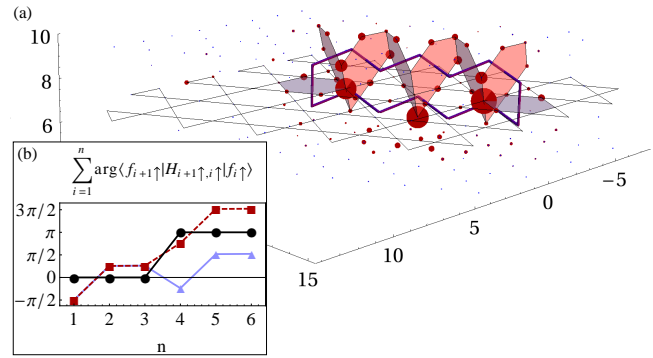}
\caption{(color online) (a) $\mathbb{Z}_2$ vortex loop (purple line) on the pyrochlore lattice, threading hexagonal (above and in the kagome plane, shaded) and triangular plaquettes (below the kagome plane, shaded). Only a single kagome plane is shown. The (red) spheres and their relative size denote the probability density of a mid-gap state in the TI* phase that is localized on the loop. The axes employ units in which nearest-neighbor displacements are $2\sqrt{2/3}$ in length. (b) Accumulated Berry phase~\cite{Levin2003} for adiabatically braiding a slave-fermion around a kagome plane hexagon pierced by the vortex loop (red dashed) and without the vortex loop (light blue). The difference of the two (solid black) yields the predicted statistical phase of $\pi$.}\label{LGT}
\end{figure}

The prediction of nontrivial mutual statistics between slave-fermions and $\mathbb{Z}_2$ vortex loops encoded in the topological field theory (\ref{LTI}) can be tested numerically. A plaquette threaded by a vortex line satisfies $\prod_{rr'} \sigma_{rr'}=-1$, where the product runs over the plaquette edges. We consider a $\sigma_{rr'}$ configuration with a single $\mathbb{Z}_2$ vortex loop in a finite periodic pyrochlore cluster [Fig.~\ref{LGT}(a)]. With uniform amplitudes of $\chi$ and $J$, $H_\textrm{MF}$ yields Kramers doublets in the bulk energy gap that are localized on the vortex loop. We find numerically [Fig.~\ref{LGT}(b)] that braiding a slave-fermion along a path linking the vortex loop yields a statistical phase of $\pi$, consistent with Eq.~(\ref{LTI}).

In conclusion, we have proposed a new possible topological phase of strongly correlated 3D TI, the TI*. Because the topology of the slave-fermion band structure requires time-reversal symmetry, this is an example of symmetry-enriched topological (SET) phase. While most studies to date have focused on bosonic SET phases~\cite{essin2013,mesaros2013,
hung2013,hung2013b,lu2013}, the TI* is an example of fermionic SET phase. Its low-energy topological field theory (\ref{LTI}) is similar to that proposed by Cho and Moore~\cite{cho2011} for \emph{noninteracting} TI based on their response to \emph{external} $\pi$ flux lines, a perturbation which preserves time-reversal symmetry $\mathcal{T}$. By contrast, here Eq.~(\ref{LTI}) describes a \emph{strongly correlated} phase with nontrivial ground-state degeneracy on $T^3$ due to the dynamical nature and compactness of the gauge potentials $a_\mu$ and $b_{\mu\nu}$, and $\pi$ flux loops arise as emergent \emph{dynamical} excitations coupled to $b_{\mu\nu}$. Microscopically, these $\pi$ flux loops are gauge-invariant vortex excitations of an emergent $\mathbb{Z}_2$ lattice gauge field.  $\Sigma^{0i}$ represents the density of vortices with ``magnetic'' gauge flux along the $i$ direction and is thus odd under $\mathcal{T}$, whereas $\epsilon^{ijk}\Sigma^{jk}$ represents the density of ``dual'' vortices which carry ``electric'' gauge flux in the $i$ direction and is even under $\mathcal{T}$. The slave-particle density $j^0$ and current $j^i$ are even and odd, respectively. Since the gauge potentials $a_\mu$ and $b_{\mu\nu}$ transform oppositely to the currents $j^\mu$ and $\Sigma^{\mu\nu}$ to which they couple, we find that $a_0$ and $b_{ij}$ are even under $\mathcal{T}$, whereas $a_i$ and $b_{0i}$ are odd. This implies that the effective theory (\ref{LTI}) preserves $\mathcal{T}$. As a first step towards demonstrating the feasibility of the TI* in a real material, we performed the slave-spin mean-field analysis on a model of interacting fermions on the pyrochlore lattice. We found the TI* and several gapless semi-metallic SM* phases in this model for a wide range of parameters. These phases do not necessarily require a very strong Hubbard interaction, but do require strong spin-orbit coupling. The phase diagram of this model contains an interesting multicritical point which tends to favor the TI* and SM* phases due to the reduced electron band gap in its vicinity. Strong pair-hopping interactions~\cite{nandkishore2012} in addition to a strong Hubbard repulsion might also favor these phases over the TMI or conventional magnetic phases.

We are grateful to A. R\"uegg for numerous discussions and collaborations on closely related projects, as well as G. Y. Cho, Y. B. Kim, X. L. Qi, R. Nandkishore, M. Schir\'{o}, and T. Senthil for useful discussions. JM acknowledges financial support from the Simons Foundation. VC and GAF acknowledge financial support through ARO Grant No. W911NF-09-1-0527, NSF Grant No. DMR-0955778, and grant W911NF-12-1-0573 from the Army Research Office with funding from the DARPA OLE Program.

\bibliography{z2ti}

\begin{thebibliography}{54}
\bibitem [{BFn()}]{BFnormalization}%
  \BibitemOpen
  \href@noop {} {}\bibinfo {note} {The factor $\frac{1}{4\pi}$ in
  Eq.~(\ref{LTI}) corresponds to a choice of normalization of the compact
  2-form gauge field $b_{\mu\nu}$ such that for periodic boundary conditions in
  all three directions of space, $b_{ij}$ is equivalent to $b_{ij}+\frac{2\pi
  m_{ij}}{L_iL_j}$ under a large 2-form gauge transformation on the 2-cycle
  generated by the orthogonal directions $i$ and $j$ ($i,j=1,2,3$), where the
  integer $m_{ij}$ is the winding number of the gauge transformation and $L_i$
  is the linear size of the system along direction $i$.}\BibitemShut {Stop}%
\end{thebibliography}

\end{document}


\renewcommand{\Re}{\mathop{\mathrm{Re}}}
\renewcommand{\Im}{\mathop{\mathrm{Im}}}
\renewcommand{\b}[1]{\mathbf{#1}}
\renewcommand{\u}{\uparrow}
\renewcommand{\d}{\downarrow}
\newcommand{\bsigma}{\boldsymbol{\sigma}}
\newcommand{\blambda}{\boldsymbol{\lambda}}
\newcommand{\Tr}{\mathop{\mathrm{Tr}}}
\newcommand{\sgn}{\mathop{\mathrm{sgn}}}
\newcommand{\sech}{\mathop{\mathrm{sech}}}
\newcommand{\diag}{\mathop{\mathrm{diag}}}
\newcommand{\half}{{\textstyle\frac{1}{2}}}
\newcommand{\sh}{{\textstyle{\frac{1}{2}}}}
\newcommand{\ish}{{\textstyle{\frac{i}{2}}}}
\newcommand{\thf}{{\textstyle{\frac{3}{2}}}}
\newcommand{\be}{\begin{equation}}
\newcommand{\ee}{\end{equation}}

\renewcommand{\thetable}{S\Roman{table}}
\renewcommand{\thefigure}{S\arabic{figure}}
\renewcommand{\thesubsection}{S\arabic{subsection}}
\renewcommand{\theequation}{S\arabic{equation}}

\title{Supplemental material for ``Topological order in a correlated three-dimensional topological insulator''}

\author{Joseph Maciejko}
\email[electronic address: ]{jmaciejk@princeton.edu}
\affiliation{Princeton Center for Theoretical Science, Princeton University, Princeton, New Jersey 08544, USA}

\author{Victor Chua}
\affiliation{Department of Physics, The University of Texas at Austin, Austin, Texas 78712, USA}

\author{Gregory A. Fiete}
\affiliation{Department of Physics, The University of Texas at Austin, Austin, Texas 78712, USA}

\date\today

\maketitle

This supplemental material contains technical details omitted in the main text and is divided in three parts. In Sec.~\ref{sec:pathintegral}, we give further details concerning the Hamiltonian formulation of the $\mathbb{Z}_2$ slave-spin theory and its path-integral representation. In Sec.~\ref{sec:U1gauge}, we give the technical details of the derivation of the topological field theory of the TI* phase [Eq.~(7) in the main text] from the $\mathbb{Z}_2$ gauge theory [Eq.~(3) in the main text]. Finally, in Sec.~\ref{sec:lattice} we supply details regarding the geometry of the pyrochlore lattice and $\mathbb{Z}_2$ vortex loops in this lattice. In addition, we provide a Mathematica notebook file (\texttt{z2ti\_supplemental.nb}) that can be used to visualize the vortex loop in 3D.

\section{Path-integral formulation of the $\mathbb{Z}_2$ slave-spin theory}
\label{sec:pathintegral}

We start with the electron lattice model of Eq.~(1) in the main text,
\begin{align}\label{H}
H=\sum_{rr'}\sum_{\alpha\beta}
t_{\alpha\beta}^{rr'}
\hat{c}_{r\alpha}^\dag\hat{c}_{r'\beta}
+\frac{U}{2}\sum_r\left(\sum_\alpha\hat{c}_{r\alpha}^\dag\hat{c}_{r\alpha}
-1\right)^2,
\end{align}
where $r,r'$ are site indices, $\alpha,\beta$ are spin indices, $U>0$ is the on-site Coulomb repulsion, and $t_{\alpha\beta}^{rr'}$ are hopping matrices. In the $\mathbb{Z}_2$ slave-spin approach,\cite{huber2009,ruegg2010,nandkishore2012,
ruegg2012,maciejko2013} we ``fractionalize'' the electron operator into a product of a slave-fermion $f$ and a Pauli matrix $\tau^x$,
\begin{align}
\hat{c}_{r\alpha}=\hat{f}_{r\alpha}\hat{\tau}^x_r,
\end{align}
where $\hat{\tau}^x_r$ acts on an Ising slave-spin. In contrast to the main text, in this supplemental material we explicitly denote quantum operators by a caret (e.g., $\hat{f}_{r\alpha},\hat{f}_{r\alpha}^\dag$), to distinguish them from quantum fields (e.g., $f_{r\alpha},\bar{f}_{r\alpha}$) in the path-integral formulation to follow. Unoccupied/doubly occupied states have slave-spin $\hat{\tau}^z_r=1$ and singly occupied states have slave-spin $\hat{\tau}^z_r=-1$. Since $(\hat{n}_r-1)^2=1$ for unoccupied/doubly occupied states and $(\hat{n}_r-1)^2=0$ for singly occupied states, where $\hat{n}_r=\sum_a \hat{c}_{r\alpha}^\dag\hat{c}_{r\alpha}$ is the total electron number on site $r$, we have the equality $(\hat{n}_r-1)^2=\half(\hat{\tau}^z_r+1)$. Furthermore, given that $(\hat{\tau}^x_r)^2=1$, the Hamiltonian (\ref{H}) can be written in terms of slave-fermions and slave-spins as
\begin{align}\label{Hslave}
H=H_t+H_U,
\end{align}
where
\begin{align}
H_t&=\sum_{rr'}\sum_{\alpha\beta}
t_{\alpha\beta}^{rr'}
\hat{\tau}^x_r\hat{\tau}^x_{r'}\hat{f}_{r\alpha}^\dag
\hat{f}_{r'\beta},\label{Ht}\\
H_U&=\frac{U}{4}\sum_r(\hat{\tau}^z_r+1).
\end{align}
This Hamiltonian is invariant under the local (gauge) $\mathbb{Z}_2$ transformations
\begin{align}\label{Z2symmetry}
\hat{f}_{r\alpha}\rightarrow\varepsilon_r \hat{f}_{r\alpha},
\hspace{5mm}
\hat{f}^\dag_{r\alpha}\rightarrow\varepsilon_r \hat{f}^\dag_{r\alpha},
\hspace{5mm}
\hat{\tau}^x_r\rightarrow\varepsilon_r\hat{\tau}^x_r,
\end{align}
where $\varepsilon_r=\pm 1$. However, this Hamiltonian acts on a Hilbert space which is bigger that the electron Hilbert space due to the presence of states that violate the local constraint $(\hat{n}_r-1)^2=\half(\hat{\tau}^z_r+1)$. This constraint is equivalent to the constraint $\hat{G}_r=1$ where we define the unitary operator
\begin{align}
\hat{G}_r\equiv(-1)^{\hat{n}_r+\half(\hat{\tau}^z_r-1)},
\end{align}
which performs a $\mathbb{Z}_2$ gauge transformation at site $r$,
\begin{align}
\hat{G}_r \hat{f}_{r\alpha} \hat{G}_r=-\hat{f}_{r\alpha},
\hspace{5mm}
\hat{G}_r \hat{f}^\dag_{r\alpha} \hat{G}_r=-\hat{f}^\dag_{r\alpha},
\hspace{5mm}
\hat{G}_r \hat{\tau}^x_r \hat{G}_r=-\hat{\tau}^x_r.
\end{align}
The physical states are the states that respect the local constraint $\hat{G}_r=1$, i.e., the states $|\psi\rangle$ such that $\hat{G}_r|\psi\rangle=|\psi\rangle$, $\forall r$. In other words, the physical states are gauge invariant. The Hamiltonian is also gauge invariant, $[\hat{G}_r,H]=0$ for all sites $r$. We can also say that the physical states satisfy $\hat{P}|\psi\rangle=|\psi\rangle$ while the unphysical states satisfy $\hat{P}|\psi\rangle=0$, where $\hat{P}$ is a projector defined as
\begin{align}\label{projector}
\hat{P}=\prod_r \hat{P}_r,\hspace{5mm}
\hat{P}_r=\frac{1}{2}(1+\hat{G}_r).
\end{align}
The original electron problem (\ref{H}) is equivalent to using the Hamiltonian (\ref{Hslave}) but working only with physical states. In other words, we should use $P$ to project out the unphysical states.

We now follow the Senthil-Fisher approach\cite{senthil2000} to derive a path-integral representation of the partition function. The main steps of the procedure are as follows. We first introduce coherent states for the degrees of freedom of the theory---slave-fermions and slave-spins---as is conventional in any path-integral representation. However, in our problem the slave-fermions and slave-spins are not free to fluctuate independently, but are tied together by the local constraint $\hat{G}_r=1$. This constraint is implemented in the path integral by introducing a new Ising variable $\sigma_{r\tau}=\pm 1$, which is essentially a Lagrange multiplier for each site $r$ of the spatial lattice. Next, we decouple the four-operator term (\ref{Ht}) using a Hubbard-Stratonovich field. After a saddle-point approximation, this Hubbard-Stratonovich field gives rise to another Ising variable $\sigma_{rr'}=\pm 1$ that lives on the links $rr'$ of the spatial lattice. Together, $\sigma_{r\tau}$ and $\sigma_{rr'}$ respectively form the temporal and spatial components of a dynamical, spacetime $\mathbb{Z}_2$ gauge field, to which the slave-fermions and slave-spins are minimally coupled.

The saddle-point approximation neglects amplitude fluctuations, and corresponds to what Wen calls \emph{first-order mean-field theory}.\cite{WenBook} As discussed by Senthil and Fisher,\cite{senthil2000} one expects this approximation to become exact in a suitable large-$N$ limit, where $N$ is the number of species of slave-spins and slave-fermions.

We now outline the procedure in greater detail. In the partition sum we should only sum over physical states, which is accomplished by inserting the projector $\hat{P}$ inside the trace,
\begin{align}
Z=\Tr (e^{-\beta H}\hat{P}).
\end{align}
Note that $\hat{P}^2=\hat{P}$ (projector) and $[\hat{P},H]=0$ since $H$ is gauge invariant. Using the Suzuki-Trotter expansion, we have
\begin{align}\label{suzukitrotter}
Z=\Tr[(e^{-\epsilon H}\hat{P})^M],
\end{align}
where $\epsilon=\beta/M\rightarrow 0$ and $M\rightarrow\infty$ is the number of imaginary time slices. We work with the slave-fermion coherent states
\begin{align}
|f\rangle&=e^{-\sum_{r\alpha}f_{r\alpha}\hat{f}^\dag_{r\alpha}}
|0\rangle,\\
\langle\bar{f}|&=\langle 0|e^{\sum_{r\alpha}\bar{f}_{r\alpha}\hat{f}_{r\alpha}},
\end{align}
while for slave spins, we work in the $\tau^x$ basis,
\begin{align}
\hat{\tau}^x_r|\tau^x\rangle=\tau^x_r|\tau^x\rangle,
\,\forall r.
\end{align}
Tracing over slave-fermions and slave-spins and using the resolution of the identity for both degrees of freedom
\begin{align}
\int\prod_{r\alpha}d\bar{f}_{r\alpha\tau}df_{r\alpha,\tau+1}
e^{-\sum_{r\alpha}\bar{f}_{r\alpha\tau}f_{r\alpha,\tau+1}}
|f_{\tau+1}\rangle\langle\bar{f}_\tau|&=1,\\
\sum_{\{\tau^x_{r\tau}\}}|\tau^x_\tau\rangle\langle\tau^x_\tau|&=1,
\end{align}
where $\tau=1,\ldots,M-1$ is the time-slice index, we obtain
\begin{align}\label{Ztrace2}
Z=\int\prod_{\tau=1}^M\prod_{r\alpha}d\bar{f}_{r\alpha\tau}
df_{r\alpha\tau}\sum_{\{\tau^x_{r\tau}\}}
e^{-\sum_{\tau=1}^M\sum_{r\alpha}\bar{f}_{r\alpha\tau}
f_{r\alpha,\tau+1}}\langle\bar{f}_\tau,\tau^x_\tau|
e^{-\epsilon H}\hat{P}|f_\tau,\tau^x_{\tau-1}\rangle,
\end{align}
with the boundary conditions in the imaginary time direction
\begin{align}\label{bc}
f_{M+1}=-f_1,\hspace{5mm}
\tau^x_0=\tau^x_M.
\end{align}
The projector (\ref{projector}) can be written as
\begin{align}
\hat{P}=\prod_r\frac{1}{2}\left[1+(-1)^{\sum_\alpha \hat{f}^\dag_{r\alpha}\hat{f}_{r\alpha}
+\half(\hat{\tau}^z_r-1)}\right]
=\prod_r\frac{1}{2}\sum_{\sigma_{r\tau}=\pm 1}
e^{(i\pi/2)(1-\sigma_{r\tau})\left[\sum_\alpha \hat{f}^\dag_{r\alpha}\hat{f}_{r\alpha}
+\half(\hat{\tau}^z_r-1)\right]},
\end{align}
where we have introduced a new Ising variable $\sigma_{r\tau}$ at each site $r$ and for each time slice $\tau$ to implement the local constraint. The matrix element $\langle\bar{f}_\tau,\tau^x_\tau|
e^{-\epsilon H}\hat{P}|f_\tau,\tau^x_{\tau-1}\rangle$ is given by
\begin{align}
\langle\bar{f}_\tau,\tau^x_\tau|
e^{-\epsilon H}\hat{P}|f_\tau,\tau^x_{\tau-1}\rangle=\sum_{\{\sigma_{r\tau}\}}
\sum_{\{\tau^z_{r\tau}\}}
e^{\sum_{r\alpha}\bar{f}_{r\alpha\tau}\sigma_{r\tau}
f_{r\alpha\tau}}e^{-\epsilon H(\tau^x_\tau,\tau^z_\tau,
\bar{f}_\tau,\sigma_\tau f_\tau)}
e^{(i\pi/4)\sum_r(1-\tau^z_{r\tau})[\tau^x_{r\tau}-\tau^x_{r,\tau-1}-(1-\sigma_{r\tau})]},
\end{align}
where we have neglected multiplicative factors of $\half$. Upon performing the change of variables $\sigma_\tau f_\tau\rightarrow f_\tau$, the partition function (\ref{Ztrace2}) becomes
\begin{align}
Z=\int\prod_{\tau=1}^M\prod_{r\alpha}d\bar{f}_{r\alpha\tau} df_{r\alpha\tau}\sum_{\{\tau^x_{r\tau}\}}
\sum_{\{\tau^z_{r\tau}\}}\sum_{\{\sigma_{r\tau}\}}e^{-S},
\end{align}
where the imaginary-time action $S$ is
\begin{align}
S=S_\tau^f+S_\tau^\textrm{Ising}
+\epsilon\sum_{\tau=1}^MH(\tau^x_\tau,\tau^z_\tau,\bar{f}_\tau,f_\tau),
\end{align}
with
\begin{align}
S_\tau^f&=\sum_{\tau=1}^M\sum_{r\alpha}
\bar{f}_{r\alpha\tau}(\sigma_{r,\tau+1}f_{r\alpha,\tau+1}
-f_{r\alpha\tau}),\\
S_\tau^\textrm{Ising}&=-\frac{i\pi}{4}\sum_{\tau=1}^M
\sum_r(1-\tau^z_{r\tau})[\tau^x_{r\tau}-\tau^x_{r,\tau-1}
-(1-\sigma_{r\tau})],
\end{align}
with the boundary conditions
\begin{align}
f_{r\alpha,\tau=M+1}=-f_{r\alpha,\tau=1},\hspace{5mm}
\tau^x_{r,\tau=M}=\tau^x_{r,\tau=0},\hspace{5mm}
\sigma_{r,\tau=M+1}=\sigma_{r,\tau=1}.
\end{align}
$\sigma_{r\tau}$ should be interpreted as the time component of a $\mathbb{Z}_2$ gauge field. Indeed, one can check that the partition function is invariant under $\mathbb{Z}_2$ transformations that are local in time,
\begin{align}
f_{r\alpha\tau}\rightarrow\varepsilon_\tau f_{r\alpha\tau},
\hspace{5mm}
\bar{f}_{r\alpha\tau}\rightarrow\varepsilon_\tau \bar{f}_{r\alpha\tau},\hspace{5mm}
\tau^x_{r\tau}\rightarrow\varepsilon_\tau\tau^x_{r\tau},
\hspace{5mm}
\sigma_{r\tau}\rightarrow\varepsilon_{\tau-1}\sigma_{r\tau}
\varepsilon_\tau,
\end{align}
where $\varepsilon_\tau=\pm 1$.

In order to obtain a full-fledged spacetime $\mathbb{Z}_2$ gauge field, we need to decouple the quartic term $H_t\sim\tau^x\tau^x\bar{f}f$ in $H$ by a Hubbard-Stratonovich transformation. This is achieved by introducing a real 2-component field $\chi_{rr'}=(\chi_{rr'}^{(1)},\chi_{rr'}^{(2)})$ which lives on the links $rr'$ of the spatial lattice. Writing $t_{\alpha\beta}^{rr'}\equiv t\Gamma_{\alpha\beta}^{rr'}$ where $t>0$ is a positive amplitude, we have
\begin{align}
e^{-\epsilon H_t}=\prod_{\tau=1}^M\prod_{rr'}
\int d\chi_{rr'}\,e^{-S_\chi},
\end{align}
where
\begin{align}\label{Schi}
S_\chi=-\epsilon\sum_{\tau=1}^M\sum_{rr'}
\left(\frac{1}{t}\chi_{rr'}^{(1)}(\tau)\chi_{rr'}^{(2)}(\tau)
+\chi_{rr'}^{(1)}(\tau)\tau^x_{r\tau}\tau^x_{r'\tau}
+\chi_{rr'}^{(2)}\sum_{\alpha\beta}\Gamma_{\alpha\beta}^{rr'}
\bar{f}_{r\alpha\tau}f_{r'\beta\tau}\right).
\end{align}
The partition function becomes
\begin{align}
Z=\int\prod_{\tau=1}^M\prod_{r\alpha}d\bar{f}_{r\alpha\tau} df_{r\alpha\tau}
\int\prod_{rr'}d\chi_{rr'}
\sum_{\{\tau^x_{r\tau}\}}
\sum_{\{\tau^z_{r\tau}\}}\sum_{\{\sigma_{r\tau}\}}e^{-S},
\end{align}
where $S=S_\tau^f+S_\tau^\textrm{Ising}+S_r$ with
\begin{align}
S_r=\epsilon\sum_{\tau=1}^MH_U+S_\chi.
\end{align}
So far, our manipulations have been exact. Now, we perform a saddle-point approximation on the path integral over $\chi_{rr'}$. We pick a uniform and constant saddle point $\chi_{rr'}^{(1)}=\chi_1$ and $\chi_{rr'}^{(2)}=\chi_2$ where $\chi_1,\chi_2$ are real constants. However, this saddle point breaks the $\mathbb{Z}_2$ symmetry (\ref{Z2symmetry}) of the original Hamiltonian. We therefore allow for sign (gauge) fluctuations in the simplest possible way,
\begin{align}
\chi_{rr'}(\tau)=\left(\begin{array}{c}
\chi_1 \\
\chi_2
\end{array}\right)\sigma_{rr'}(\tau),
\end{align}
where $\sigma_{rr'}(\tau)=\pm 1$, and we ignore amplitude fluctuations (fluctuations of $\chi_1,\chi_2$) that are massive at the saddle-point. The first term in Eq.~(\ref{Schi}) becomes a constant that we neglect. Denoting spacetime ``lattice sites'' by $i,j$, we now have a full-fledged spacetime $\mathbb{Z}_2$ gauge field $\sigma_{ij}=(\sigma_{r\tau},\sigma_{rr'})$ that lives on the links of the spacetime lattice. The partition function becomes
\begin{align}
Z=\int\prod_{i\alpha}d\bar{f}_{i\alpha}df_{i\alpha}
\sum_{\{\tau^x_i\}}\sum_{\{\tau^z_i\}}\prod_{ij}\sum_{\sigma_{ij}=\pm 1}e^{-S},
\end{align}
where $S=S_\tau^f+S_\tau^\textrm{Ising}+S_0+S_U$, with
\begin{align}
S_\tau^f&=\sum_{i,j=i+\hat{\tau}}\sum_\alpha
\bar{f}_{i\alpha}(\sigma_{ij}f_{j\alpha}-f_{i\alpha}),\\
S_\tau^\textrm{Ising}&=-\frac{i\pi}{4}\sum_{i,j=i-\hat{\tau}}
(1-\tau^z_i)[\tau^x_i-\tau^x_j-(1-\sigma_{ij})],\\
S_0&=-\epsilon\sum_{i,j=i+\hat{r}}\left(\chi_1\tau^x_i
\sigma_{ij}\tau^x_j+\chi_2\sum_{\alpha}
\Gamma_{\alpha\beta}^{ij}\bar{f}_{i\alpha}\sigma_{ij}
f_{j\beta}\right),\\
S_U&=\frac{\epsilon U}{4}\sum_i(\tau^z_i+1),
\end{align}
where we denote $\sigma_{i,i-\hat{\tau}}\equiv\sigma_{r\tau}$. To obtain an effective action solely in terms of slave-fermions $f,\bar{f}$, slave-spins $\tau^x$, and $\mathbb{Z}_2$ gauge fields $\sigma_{ij}$, we need to perform the sum over $\tau^z_i$, which involves $S_\tau^\textrm{Ising}$ and $S_U$. Neglecting constant multiplicative factors, we have
\begin{align}\label{SuSIsing}
\sum_{\{\tau^z_i\}}e^{-(S_U+S_\tau^\textrm{Ising})}
=e^{\half\ln\coth(\epsilon U/2)\sum_i\tau^x_i\sigma_{i,i-\hat{\tau}}
\tau^x_{i-\hat{\tau}}}e^{-S_B},
\end{align}
where $S_B$ is a Berry phase term\cite{senthil2000} given by
\begin{align}
e^{-S_B}=\prod_{i,j=i-\hat{\tau}}\sigma_{ij},
\end{align}
where we have used the periodic boundary condition (\ref{bc}) on $\tau^x$. The slave-spin part of the action now reads
\begin{align}
S_{\tau^x}=-\epsilon\chi_1\sum_{i,j=i+\hat{r}}\tau^x_i
\sigma_{ij}\tau^x_j-\half\ln\coth\left(\frac{\epsilon U}{2}\right)
\sum_{i,j=i-\hat{\tau}}\tau^x_i\sigma_{ij}\tau^x_j.
\end{align}
Following Ref.~\onlinecite{senthil2000}, we make the hopping in the space and time directions the same by a special choice of $\epsilon$ which corresponds to a special choice of ultraviolet regularization. Holding $\chi_1>0$ fixed, the equation $\epsilon\chi_1=\half\ln\coth(\epsilon U/2)$ can be solved numerically for $\epsilon$ as a function of $U$. Defining $\kappa\equiv\epsilon\chi_1$, we find that $\kappa(U)$ is a positive and monotonically decreasing function of $U$ which has the following limits,
\begin{align}\label{kappalimits}
\lim_{U\rightarrow 0}\kappa(U)=\infty,\hspace{5mm}
\lim_{U\rightarrow\infty}\kappa(U)=0.
\end{align}
We therefore arrive at Eq.~(3) of the main text, the partition function of a 4D Euclidean $\mathbb{Z}_2$ gauge theory with bosonic and fermionic matter in the fundamental representation,
\begin{align}
Z=\int D\bar{f}_{i\alpha}Df_{i\alpha}
\sum_{\{\tau^x_i\}}\sum_{\{\sigma_{ij}\}}
e^{-S_{\mathbb{Z}_2}[\bar{f},f,\tau^x,\sigma]},
\end{align}
where the action $S_{\mathbb{Z}_2}=S_{\tau^x}+S_f+S_B$ is
\begin{align}
S_{\tau^x}&=-\kappa\sum_{ij}\tau^x_i
\sigma_{ij}\tau^x_j,\\
S_f&=-\sum_{ij}
\sum_{\alpha\beta}t^{ij}_{\alpha\beta}
\bar{f}_{i\alpha}\sigma_{ij}f_{j\beta},\label{Sfsigmaij}\\
e^{-S_B}&=\prod_{i,j=i-\hat{\tau}}\sigma_{ij},
\end{align}
with $t^{ij}_{\alpha\beta}$ equal to $\epsilon\chi_2\Gamma^{ij}_{\alpha\beta}$ on spatial nearest-neighbor links and $-\delta_{\alpha\beta}$ on  temporal nearest-neighbor links. In the next subsections we consider two limiting cases: $U=0$ and $U=\infty$.

\subsection{$U=0$ limit: topological band insulator}

In the $U=0$ limit, Eq.~(\ref{SuSIsing}) becomes
\begin{align}
\sum_{\{\tau^z_i\}}e^{-(S_U+S_\tau^\textrm{Ising})}
=\prod_i \left(1
+\tau^x_i\sigma_{i,i-\hat{\tau}}\tau^x_{i-\hat{\tau}}\right),
\end{align}
which kills the path integral unless $\tau^x_i\sigma_{i,i-\hat{\tau}}\tau^x_{i-\hat{\tau}}=1$ on each site $i$. In particular, the product for all sites is also one,
\begin{align}
\prod_i\tau^x_i\sigma_{i,i-\hat{\tau}}\tau^x_{i-\hat{\tau}}
=\prod_{i,j=i-\hat{\tau}}\sigma_{ij}=1,
\end{align}
which can be satisfied by the choice of gauge $\sigma_{i,i-\hat{\tau}}=1$ on each temporal link. Therefore the temporal gauge fields are frozen in the $U=0$ limit. Since $\tau^x_i\sigma_{i,i-\hat{\tau}}\tau^x_{i-\hat{\tau}}=1$, this choice of gauge also implies $\tau^x_i\tau^x_{i-\hat{\tau}}=1$ on each site, which means one of two possibilities: either $\tau^x_i=1$ on each site, or $\tau^x_i=-1$. In other words, the slave-spins $\tau^x$ condense (they are ferromagnetically ordered). The electron and slave-fermion operators become proportional, $\hat{c}_{r\alpha}=\hat{f}_{r\alpha}$ or $\hat{c}_{r\alpha}=-\hat{f}_{r\alpha}$, and we recover the topological band insulator.

\subsection{$U=\infty$ limit: effective spin model}

Recall from Eq.~(\ref{kappalimits}) that $\kappa=0$ in this limit, hence $S_{\tau^x}=0$ and the slave-spins $\tau^x$ can be trivially integrated out. We can perform the trace over $\sigma_{ij}$ on spatial links which only involves $S_f$ [Eq.~(\ref{Sfsigmaij})], to obtain an effective action on the spatial links
\begin{align}\label{largeU4fermion}
\tilde{S}_f^r=-\sum_{rr'}\sum_\tau\ln\cosh\left(
\sum_{\alpha\beta} t_{\alpha\beta}^{rr'}\bar{f}_{r\alpha\tau}
f_{r'\beta\tau}\right)
=-\frac{1}{2}\sum_{rr'}\sum_\tau
\sum_{\alpha\beta\gamma\delta}t_{\alpha\beta}^{rr'}
t_{\gamma\delta}^{rr'}\bar{f}_{r\alpha\tau}
f_{r'\beta\tau}\bar{f}_{r\gamma\tau}f_{r'\delta\tau}+\ldots,
\end{align}
where the extra terms involve at least eight fermion operators. We have generated four-fermion interactions. Performing the trace over $\sigma_{ij}$ on temporal links which involves $S_f$ as well as the Berry phase term $S_B$, we find the constraint
\begin{align}
(-1)^{\hat{n}_r}=-1,
\end{align}
on each site $r$. In other words, the limit $U\rightarrow\infty$ is described by the four-fermion interaction (\ref{largeU4fermion}) with the constraint of one fermion per site. Therefore we have a Heisenberg-type spin-$\frac{1}{2}$ model,
\begin{align}\label{HeffUlarge}
H_\textrm{eff}(U\rightarrow\infty)=\sum_{rr'}
J^{\mu\nu}_{rr'}S^\mu_rS^\nu_{r'}+\ldots,
\end{align}
where the magnitude of $J^{\mu\nu}_{rr'}$ is of order $\sim\epsilon(\chi_2\Gamma)^2$ and the dependence on $rr'$ and $\mu\nu$ can be extracted from $t_{\alpha\beta}^{rr'}
t_{\gamma\delta}^{rr'}$. It is most likely that Eq.~(\ref{HeffUlarge}) has a magnetic ground state. (A spin liquid ground state is an another possibility.)

\section{Derivation of the topological field theory of the TI* phase}
\label{sec:U1gauge}

In this section, we provide technical details of the derivation of the topological field theory of the TI* phase [Eq. (7) in the main text]. While the $\theta$-term can be expected in the topological field theory of the TI* because the slave-fermions have a topological band structure, the derivation of the $BF$ term is more subtle. The presence of a level-2 $BF$ term in the topological field theory of a $\mathbb{Z}_2$ gauge theory is a specific instance of the more general fact that $\mathbb{Z}_p$ gauge theories are described by level-$p$ $BF$ theories.\cite{banks2011,maldacena2001} As mentioned in the main text, a $\mathbb{Z}_2$ gauge theory can be written as a $U(1)$ gauge theory coupled to an additional charge-2 scalar field that Bose condenses. More formally, via the mapping $\sigma_{ij}=e^{ia_{ij}}$ where $\sigma_{ij}=\pm 1$ is the $\mathbb{Z}_2$ gauge field, any $\mathbb{Z}_2$ gauge theory can be rewritten exactly using the Poisson summation formula as the theory of a compact $U(1)$ gauge field $a_{ij}\in(-\pi,\pi]$ coupled to a charge-2 integer-valued link variable $n_{ij}$ with no dynamics,\cite{ukawa1980}
\begin{align}
\prod_{ij}\sum_{\sigma_{ij}=\pm 1}
e^{-S[\sigma_{ij}]}=\prod_{ij}\int_{-\pi}^\pi
da_{ij}\sum_{n_{ij}=-\infty}^\infty e^{ip\sum_{ijs}n_{ij}a_{ij}}\exp\left(-S[\sigma_{ij}=e^{ia_{ij}}]\right),
\end{align}
where $p=2$. Applying this to Eq.~(3)-(6) in the main text, the partition function becomes
\begin{align}
Z=\int D\bar{f}_{i\alpha}Df_{i\alpha}Da_{ij}
\sum_{\{\tau^x_i\}}\sum_{\{n_{ij}\}}e^{-S_{U(1)}[\bar{f},f,\tau^x,a,A,n]},
\end{align}
where $S_{U(1)}=S_{\tau^x}+S_f+S_n+S_B$ with
\begin{align}
S_{\tau^x}&=-\kappa\sum_{ij}\tau^x_i
e^{ia_{ij}}\tau^x_j,\label{Sbeta}\\
S_f&=-\sum_{ij}
\sum_{\alpha\beta}t^{ij}_{\alpha\beta}
\bar{f}_{i\alpha}e^{i(a_{ij}+eA_{ij})}f_{j\beta},\\
S_n&=-ip\sum_{ij}n_{ij}a_{ij},\\
e^{-S_B}&=\prod_{i,j=i-\hat{\tau}}e^{ia_{ij}}.
\end{align}
Because the action is periodic $S_{U(1)}[a_{ij}]=S_{U(1)}[a_{ij}+2\pi]$, we can extend the integration over $a_{ij}$ to the real axis. For simplicity, we will assume that all links $ij$ involved in the action $S_{U(1)}$ are nearest-neighbor links, and we will use the notation $a_{i,\mu}\equiv a_{i,i+\hat{\mu}}$ and similarly for $n_{ij}$, where $\mu=0,1,2,3$ denotes spacetime directions. We have also included the external electromagnetic field $A_{ij}$, to which only the slave-fermions couple.

We can perform the shift $a_{i,\mu}\rightarrow a_{i,\mu}-\Delta_\mu\phi_i$ in the action and integrate over the real scalar field $\phi_i$, which simply overcounts the partition function by a constant multiplicative factor that leaves all physical quantities unaffected. Owing to the $U(1)$ gauge invariance of $S_{\tau^x}$, $S_f$, and $S_B$, only $S_n$ is affected by the transformation,
\begin{align}
S_n\rightarrow S_n+ip\sum_{i,\mu}n_{i,\mu}\Delta_\mu\phi_i=S_n-ip\sum_{i,\mu}\phi_i\Delta_\mu n_{i,\mu},
\end{align}
by integration by parts, where $\Delta_\mu\phi_i\equiv\phi_{i+\hat{\mu}}-\phi_i$ is the lattice derivative. Integrating over $\phi_i$, we find that $\Delta_\mu n_{i,\mu}=0$, i.e., $n_{i,\mu}$ is a conserved current. This current conservation constraint can be implemented by introducing a 2-form potential $b_{i,\mu\nu}$,
\begin{align}
n_{i,\mu}=\frac{1}{4\pi}\epsilon_{\mu\nu\lambda\rho}\Delta_\nu b_{i,\lambda\rho},
\end{align}
which, since $n_{i,\mu}\in\mathbb{Z}$, implies that $b_{i,\mu\nu}\in 2\pi\mathbb{Z}$. The partition function becomes
\begin{align}
Z=\sum_{\{\tau^x_i\}}\int D\bar{f}_{i\alpha}Df_{i\alpha}\int_{-\infty}^\infty Da_{i,\mu}\sum_{b_{i,\mu\nu}\in 2\pi\mathbb{Z}}e^{-(S_{\tau^x}+S_f+S_B)}
\exp\left(-\frac{ip}{4\pi}\sum_i\epsilon_{\mu\nu\lambda\rho}b_{i,\mu\nu}\Delta_\lambda a_{i,\rho}\right).
\end{align}
The sum over $b_{i,\mu\nu}$ is only over its six independent components,
\begin{align}
b_{i,01}=-b_{i,10},\hspace{5mm}
b_{i,02}=-b_{i,20},\hspace{5mm}
b_{i,03}=-b_{i,30},\hspace{5mm}
b_{i,12}=-b_{i,21},\hspace{5mm}
b_{i,23}=-b_{i,32},\hspace{5mm}
b_{i,31}=-b_{i,13}.
\end{align}
The integer constraint $b_{i,\mu\nu}\in 2\pi\mathbb{Z}$ can be imposed by introducing another 2-form field $\Sigma_{i,\mu\nu}\in\mathbb{Z}$,
\begin{align}
Z=\sum_{\{\tau^x_i\}}\int D\bar{f}_{i\alpha}Df_{i\alpha}\int_{-\infty}^\infty Da_{i,\mu}\sum_{\Sigma_{i,\mu\nu}\in\mathbb{Z}}
\int_{-\infty}^\infty Db_{i,\mu\nu}
e^{-(S_{\tau^x}+S_f+S_B)}
\exp\sum_i\left(-\frac{ip}{4\pi}\epsilon_{\mu\nu\lambda\rho}b_{i,\mu\nu}\Delta_\lambda a_{i,\rho}-\frac{i}{2}\Sigma_{i,\mu\nu}b_{i,\mu\nu}\right),
\end{align}
where the $\frac{1}{2}$ in front of $\Sigma_{i,\mu\nu}b_{i,\mu\nu}$ is to avoid overcounting, since $\Sigma_{i,\mu\nu}$ also only has six independent components. That this procedure implements the integer constraint can be seen by using the Poisson summation formula,
\begin{align}\label{integerconstraint}
\sum_{\Sigma_{i,\mu\nu}\in\mathbb{Z}}\exp\sum_i\left(-\frac{i}{2}\Sigma_{i,\mu\nu}b_{i,\mu\nu}\right)\propto\prod_{i,\mu<\nu}\left(\sum_{s_{i,\mu\nu}\in\mathbb{Z}}\delta(b_{i,\mu\nu}-2\pi s_{i,\mu\nu})\right).
\end{align}
As mentioned in the main text, the field $\Sigma_{i,\mu\nu}$ can be understood as a density of vortex loops that couple to the 2-form gauge potential $b_{i,\mu\nu}$.

Instead of imposing the hard integer constraint $b_{i,\mu\nu}\in 2\pi\mathbb{Z}$ exactly, it is convenient to soften it\cite{fisher1989} by introducing a small vortex core energy term $\propto\Sigma_{i,\mu\nu}^2$. We therefore write
\begin{align}\label{softconstraint}
Z=\sum_{\{\tau^x_i\}}\int D\bar{f}_{i\alpha}Df_{i\alpha}\int_{-\infty}^\infty Da_{i,\mu}&\sum_{\Sigma_{i,\mu\nu}\in\mathbb{Z}}
\int_{-\infty}^\infty Db_{i,\mu\nu}
e^{-(S_{\tau^x}+S_f+S_B)}\nonumber\\
&\times\exp\sum_i\left(-\frac{ip}{4\pi}\epsilon_{\mu\nu\lambda\rho}b_{i,\mu\nu}\Delta_\lambda a_{i,\rho}
-\frac{\kappa}{4}\Sigma_{i,\mu\nu}^2
-\frac{i}{2}\Sigma_{i,\mu\nu}b_{i,\mu\nu}\right),
\end{align}
where $\kappa\ll 1$. Mathematically, this procedure converts the delta functions in Eq.~(\ref{integerconstraint}) into narrow Gaussians of width $\propto\sqrt{\kappa}$.

We seek the topological field theory of the TI* phase, which describes its ground state. The ground state of the TI* phase corresponds to energies much less than the vortex core energy, i.e., energies such that there are no vortex loop excitations and only terms with $\Sigma_{i,\mu\nu}=0$ contribute to the partition function. We therefore have
\begin{align}
Z_\textrm{TI*}\simeq\sum_{\{\tau^x_i\}}\int D\bar{f}_{i\alpha}Df_{i\alpha}\int_{-\infty}^\infty Da_{i,\mu}\int_{-\infty}^\infty Db_{i,\mu\nu}
e^{-(S_{\tau^x}+S_f+S_B)}\exp\sum_i\left(-\frac{ip}{4\pi}\epsilon_{\mu\nu\lambda\rho}b_{i,\mu\nu}\Delta_\lambda a_{i,\rho}\right),
\end{align}
which in the continuum limit corresponds to a $BF$ term. Integrating out the slave-spins does not contribute any topological terms, while integrating out the slave-fermions gives a $\theta$-term for the combination $a_\mu+eA_\mu$ of internal and electromagnetic fields, such the topological field theory of the TI* phase in the continuum and in real time reads
\begin{align}
\mathcal{L}_\textrm{TI*}=\frac{p}{4\pi}\epsilon^{\mu\nu\lambda\rho}b_{\mu\nu}
\partial_\lambda a_\rho+\frac{\theta}{32\pi^2}\epsilon^{\mu\nu\lambda\rho}(f_{\mu\nu}+eF_{\mu\nu})(f_{\lambda\rho}+eF_{\lambda\rho}),
\end{align}
where $f_{\mu\nu}=\partial_\mu a_\nu-\partial_\nu a_\mu$ and $F_{\mu\nu}=\partial_\mu A_\nu-\partial_\nu A_\mu$ are the internal (emergent) and external (electromagnetic) field strengths, respectively. Upon performing the shift $a_\mu\rightarrow a_\mu-eA_\mu$ in the path integral, we recover Eq.~(7) of the main text.

\begin{figure}
\includegraphics[scale=0.5]{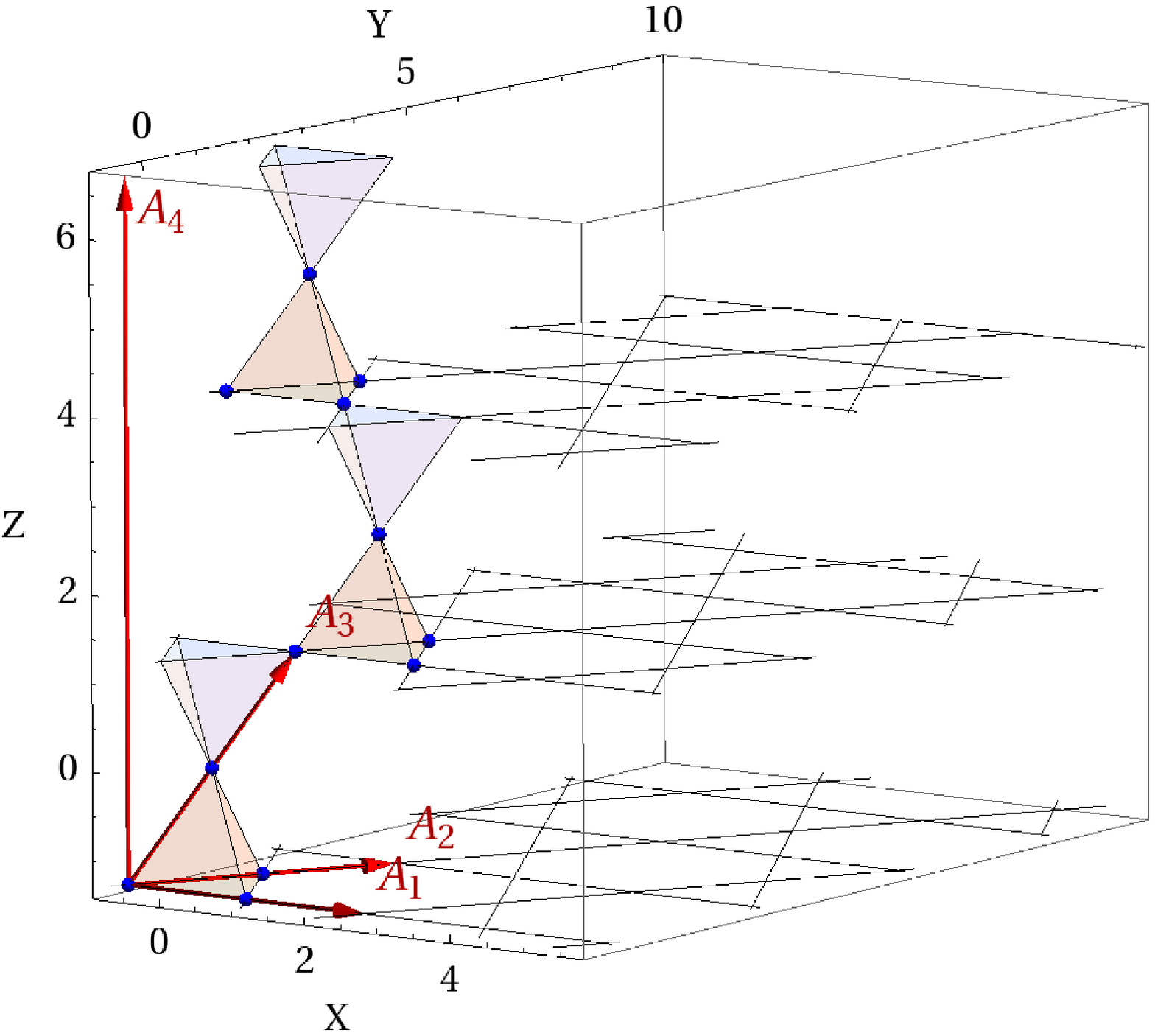}
\caption{(color online) The pyrochlore lattice with 12 sites (blue spheres) in a unit cell, composed of stacked tetrahedra (shaded). The vectors denote displacement vectors that define the geometry. Also shown are 3 parallel kagome planes formed by repeating the unit cell in the $A_1$ and $A_2$ directions.}\label{Fig3}
\end{figure}
\begin{figure}
\includegraphics[scale=0.4]{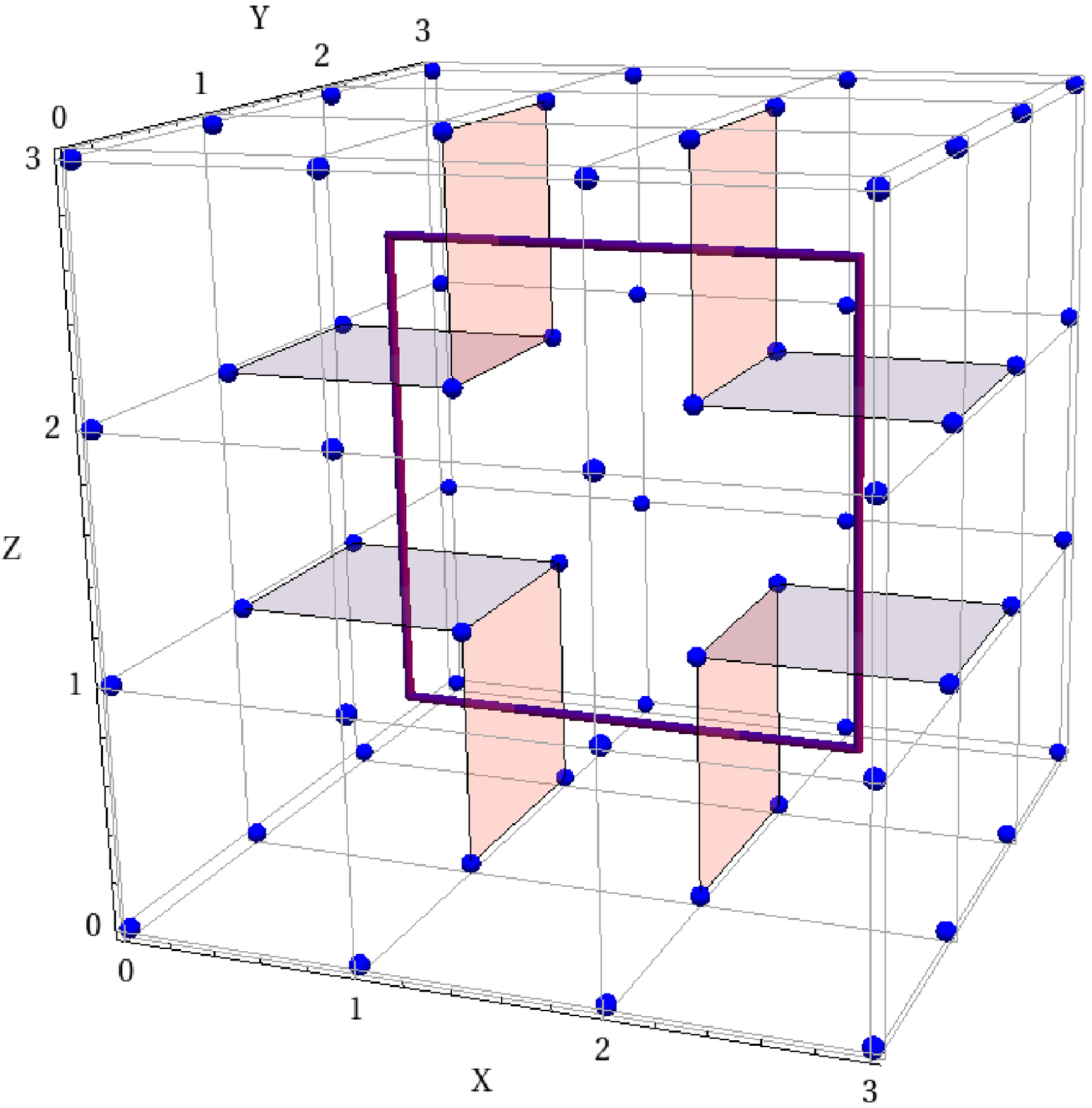}
\caption{(color online) A square vortex loop (thick purple line) in a simple cubic lattice oriented parallel to the $xz$ plane. The lattice sites are denoted by (blue) spheres and plaquettes with nonzero flux are shaded.}
\label{Fig4}
\end{figure}
\begin{figure}
\includegraphics[scale=0.5]{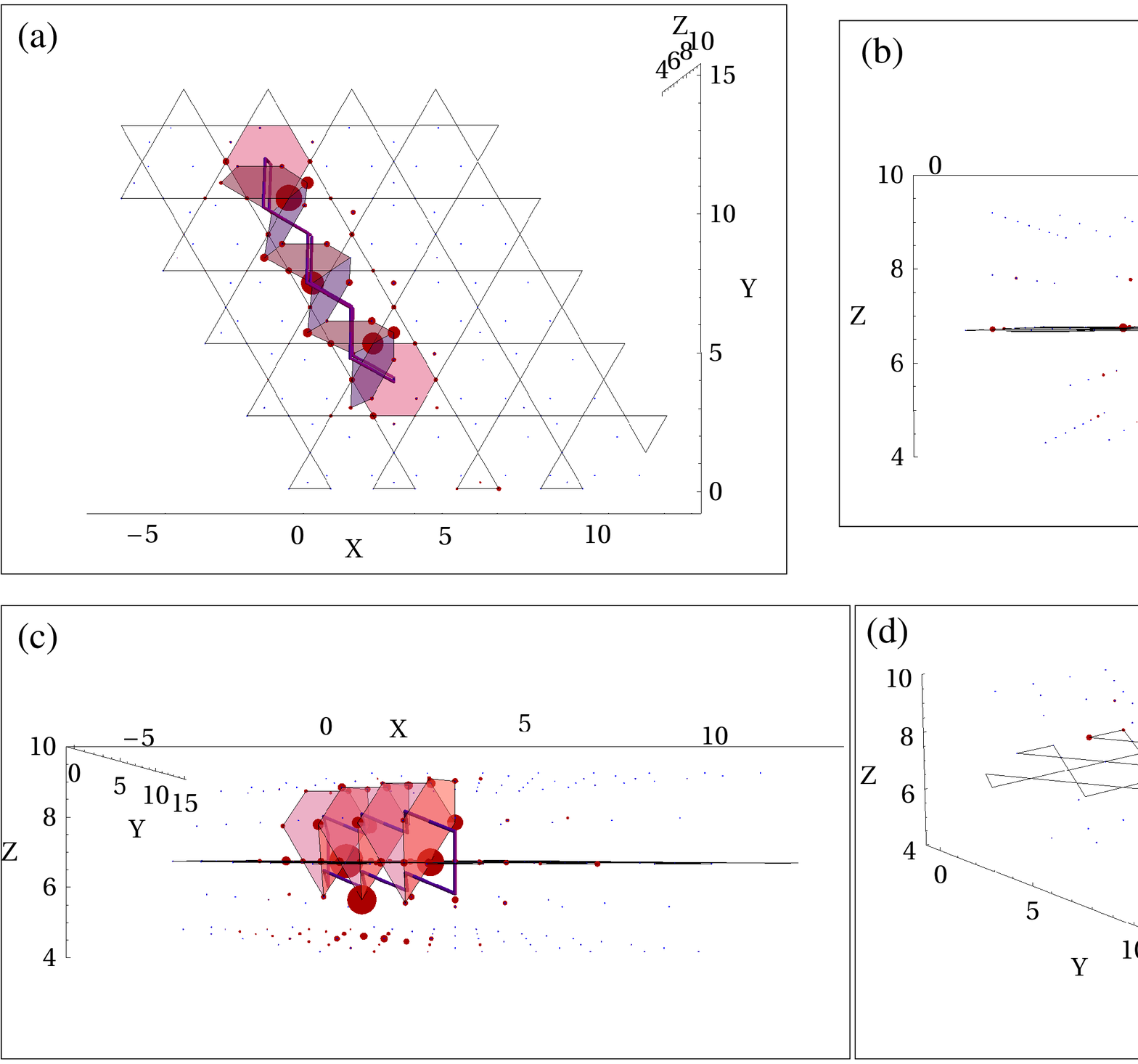}
\caption{(color online) $\mathbb{Z}_2$ vortex loop (purple line) on the pyrochlore lattice shown from different perspectives: (a) top, (b) right, and (c) front. Panel~(d) is from the same viewing angle as Fig.~2(a) of the main text. The vortex loop threads hexagonal (above and in the kagome plane, shaded) and triangular plaquettes (below the kagome plane, shaded). Only a single kagome plane is shown. The (red) spheres and their relative size denote the probability density of a mid-gap state in the TI* phase that is localized on the vortex loop.}
\label{Fig5}
\end{figure}

\section{Geometry of the pyrochlore lattice and $\mathbb{Z}_2$ vortex loops}
\label{sec:lattice}

The pyrochlore lattice is a face-centered cubic lattice composed of corner-sharing tetrahedra, with sites located at each corners. The basis of a minimal unit cell requires 4 local sites, which are related by symmetry. For our purpose of studying $\mathbb{Z}_2$ vortex loops, we have found it more convenient to use a 12-site basis (24 orbitals with spin-1/2 degeneracy) or 3 tetrahedra stacked in the $(1,1,1)$ direction relative to the conventional cubic crystallographic axes. This 12-site unit cell is shown in Fig.~\ref{Fig3} which, when repeated in the horizontal $xy$ plane, forms 3 parallel kagome planes. In our choice of length units, the relevant displacement vectors shown in Fig.~\ref{Fig3} are
\begin{eqnarray}
&&A_1=\left(4\sqrt\frac{2}{3},0,0\right), \quad A_2=\left(2 \sqrt{\frac{2}{3}},2 \sqrt{2},0\right), \quad A_3 =\left(2 \sqrt{\frac{2}{3}},\frac{2 \sqrt{2}}{3},\frac{8}{3}\right),\quad \text{and} \quad A_4=(0,0,8).
\end{eqnarray} 

In a $\mathbb{Z}_2$ lattice gauge theory, $\mathbb{Z}_2$ fluxes thread elementary plaquettes and a concatenation of them forms a closed string or vortex loop. In our visual notation, plaquettes with non-trivial flux $\prod_{rr'}\sigma_{rr'}=-1$ are shaded and pierced by a line that denotes a section of the vortex loop. As an example, we demonstrate in Fig.~\ref{Fig4} the case of a vortex loop in the a simple cubic lattice. There the only elementary plaquettes are squares that may lie parallel to the $xy$, $yz$ and $xz$ planes. 
 
By contrast, in the pyrochlore lattice, triangles and hexagons make up the set of elementary plaquettes. Moreover, they do not always lie in a kagome plane parallel to our choice of axes and may orient out of the plane. In Fig.~\ref{Fig5}, we display an example of a vortex loop in the pyrochlore lattice together with the particle densities of a midgap state in the TI* phase localized on this vortex loop. The geometry is presented from several viewing angles for clarity and complements Fig.~2(a) of the main text. We encourage the reader to use the Mathematica file (\texttt{z2ti\_supplemental.nb}) that accompanies this document to visualize the vortex loop in the pyrochlore lattice in 3D.

\bibliography{z2ti}